\documentstyle[aps,multicol,graphicx,amsmath]{revtex}

\begin{document}

\date{\today}
\draft

\title{Monitoring Single Scattering Events in Single Quantum Dots}

\author{Ulrich Hohenester$^*$}

\address{
Istituto Nazionale per la Fisica della Materia (INFM) and
Dipartimento di Fisica\\
Universit\`a di Modena e Reggio Emilia, 
Via Campi 213/A, 41100 Modena, Italy}

\maketitle

\begin{abstract} 

The quantum-jump approach is used for a theoretical description of
resonance luminescence from a {\em single}\/ semiconductor quantum dot
in contact with its solid-state environment. For continuous excitation
of the single-exciton groundstate the luminescence exhibits bright
periods, where photons are spontaneously emitted from the exciton
decay, which are interrupted by dark periods when one electron or hole
suffers a spin flip. It is shown that continuously monitored resonance
luminescence provides a very sensitive measure of such rare single
scattering events in quantum dots.

\end{abstract}



\begin{multicols}{2}

Semiconductor quantum dots (QDs) consist of a small island of
lower-bandgap material embedded in a solid-state matrix of
higher-bandgap material \cite{dot}. Proper choice of the material and
dot parameters thus can give rise to confinement of a few carrier
states within this lower-bandgap region. Because of the resulting
discrete atomic-like density-of-states, semiconductor QDs have been
referred to as {\em artificial atoms}.\/ When optical experiments are
performed on large QD ensembles, all the available samples still suffer
from the effects of inhomogeneity and dispersion in dot size. A major
advancement in the field has come from different types of local optical
experiments, that allow the investigation of individual QDs: Extremely
narrow linewidths have been reported from single-dot spectroscopy
\cite{zrenner:00}, and first time-resolved measurements indicate
carrier lifetimes a few orders of magnitude larger than typically
observed in semiconductors of higher dimension \cite{bonadeo:98}.

Theoretically the system {\em dot interacting with environment}\/ is
conveniently described within a density-matrix description
\cite{density-matrix}. Adopting the approximation of Fermi's golden
rule, at each instant of time the dot system is then completely
determined by its density matrix $\rho$ (with the diagonal and
off-diagonal elements providing information about state occupation and
coherence, respectively), and environment interactions are accounted
for by instantaneous scattering events. Because of its statistical
nature, the elements of $\rho$ have to be interpreted either as the
ensemble average over a large number of identical dots, or, for ergodic
systems, as the average over a sufficiently long measurement time.
Apparently, when performing time-resolved measurements on single dots
the situation is substantially different. Thus, the most interesting
questions arise: Does the interpretation of optical experiments of {\em
single}\/ dots require theoretical concepts beyond the usual
density-matrix description for the carrier dynamics in semiconductor
nanostructures \cite{density-matrix}, and can time-resolved single-dot
measurements provide additional information as opposed to ensemble-dot
measurements?

As we will show in the following, the answer to both questions is:
Yes. It is worth noting that related questions first arose almost a
decade ago, when it became possible to store single ions in a Paul trap
and to continuously monitor their resonance fluorescence (see Ref.
\cite{plenio:98} for a review). In the seminal work of Dalibard {\em et
al.}\/ \cite{dalibard:92}, the authors showed for a V-scheme, where the
groundstate $0$ of an atom is coupled to a short-lived state $1$ and to
a metastable state $2$, that the fluorescence of the laser-driven
system exhibits long dark periods, associated to the excitation of the
extremely weak $0\leftrightarrow 2$ transition, which are followed by
bright periods with many photon emissions from the decay of the
short-lived state $1$.

In this paper, we propose to use continuous laser excitation and
continuous monitoring of luminescence for the observation of {\em
single}\/ spin-flip processes in semiconductor QDs.  Contrary to the
proposal of Dalibard {\em et al.},\/ within our scheme the rare
scattering events originate from the coupling of the QD carriers to the
elementary excitations of their solid-state environment. Thus, the dot
serves as a {\em sensor}\/ of its environment, which makes possible the
observation of the otherwise almost inaccessible rare scattering
events. Besides the most challenging prospect of measuring single
scatterings in solid state \cite{blinking}, the long-lived spin
excitations in QDs have recently attracted strong interest
\cite{gupta:99,kamada:99,khaetskii:00} because of their potential
utilization for quantum-information processing
\cite{loss:98,imamoglu:99}.

Figure 1 sketches the proposed scheme: A single QD is located inside an
ideal broadband counter with unit efficiency and coverage of all angles
\cite{remark.detector} (grey circle). Initially, all valence-band
states of the dot are occupied and all conduction-band states are
empty. When the pump laser $\omega_p$ is turned on, an electron is
promoted from the valence to the conduction band, where the
photogenerated electron and hole (i.e., the missing valence-band
electron) have opposite spin orientations. Because of the coherence of
the driving laser the system starts do undergo Rabi-type oscillations,
i.e., the electron is transferred back and forth between the valence
and conduction band. These oscillations are interrupted by spontaneous
emissions of photons $\gamma$ from the decay of electron-hole (exciton)
states $X_{o,b}^\pm$, where, after detection of $\gamma$, the system is
reset back to the vacuum state, and the Rabi-type oscillations start
again.  After some time, however, due to environment coupling the spin
of either the electron or hole is flipped (see $\otimes$ in Fig.~1);
since each QD state can be occupied at most by two carriers with {\em
opposite}\/ spin orientations, after such a spin flip the system
becomes optically inactive (i.e., no further electron-hole pair can be
excited by the pump laser). This can be also inferred from Fig.~2(a)
which shows results of a simulation of resonance luminescence from a
single QD (for details see further below): After turning on the pump
pulse at time $t=0$, photons are emitted by spontaneous decay of
$X_{o,b}^\pm$ and are detected as single photon counts. These bright
periods are then interrupted by dark periods with zero photon count,
which indicate the occurrence of a spin flip process. It finally
requires a second spin flip to bring back the system to the bright
sector, and to turn on luminescence again.

\begin{figure}
\includegraphics[width=0.75\columnwidth]{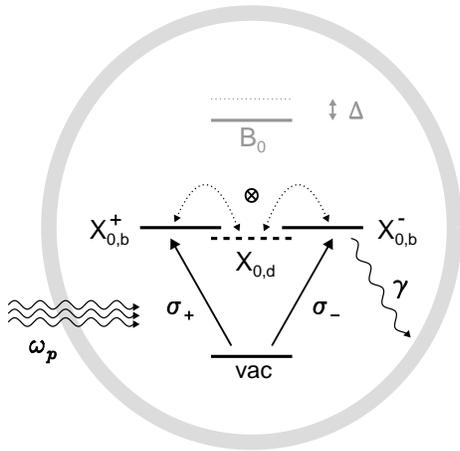}
\caption{
Schematic illustration of the proposed experimental setup, where a
single QD is surrounded by a 4$\pi$ photon detector (with unit
efficiency) which detects all spontaneously emitted photons $\gamma$.
The linearly polarized pump pulse ($\omega_p$) couples the vacuum state
to the optically allowed groundstate excitons $X_{o,b}^\pm$; $B_o$
denotes the biexciton groundstate, whose energy is by an amount of
$\Delta$ smaller than $2 E_{X_{o,b}^\pm}$ due to correlation effects.
Spin-flip processes (indicated by $\otimes$) couple the optically
allowed states $X_{o,b}^\pm$ with the optically forbidden $X_{o,d}$ 
ones, which, because of the electron-hole exchange interaction, have
a slightly smaller energy.
}
\end{figure}

The theoretical framework accounting for the coherent laser excitation
and the detection of single photon emissions is conveniently provided
by the {\em quantum-jump approach}\/ \cite{plenio:98}. Here, the
central idea is that the broad-band photon counting can be
approximately described by a series of repeated {\em gedanken}\/
measurements at times $t_n=n\cdot\Delta_t$ ($n=0,1,2,\dots$)
\cite{hegerfeldt:93}, where as observable the projector on the photon
vacuum $I\!\!P_0=|0_\gamma\rangle{\bf 1}_{\text{\tiny dot}}\langle
0_\gamma|$ indicates whether a photon has been emitted or not. If at
time $t_o$ the system is described by the density matrix $\rho_o$, the
density matrix at time $t_n$ {\em provided that no photons have been
emitted in the interval $[t_o,t_n)$}\/ is given by
\cite{plenio:98,hegerfeldt:93}:

\begin{figure}
\centerline{\includegraphics[width=0.75\columnwidth]{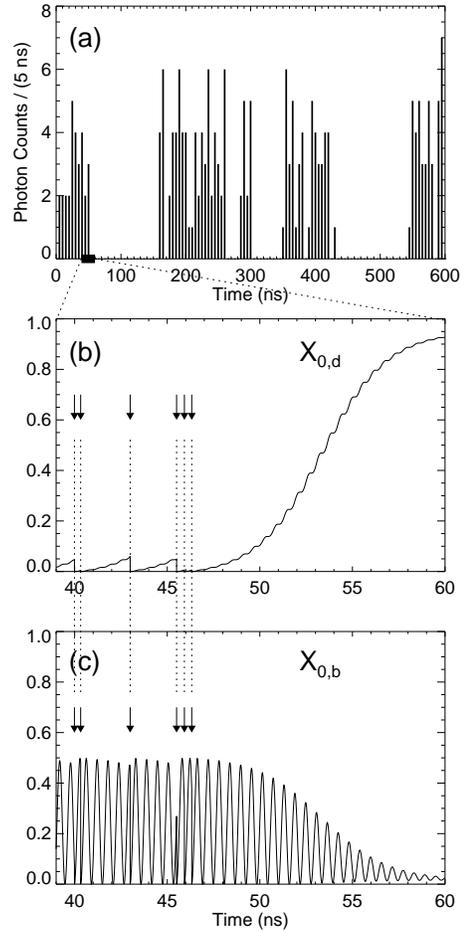}}
\caption{
Results of our Monte-Carlo simulations for resonance luminescence from
a single QD:  (a) histogram of photon emission (with bin width of of 5
ns); (b,c) trajectory of $\rho_{xx}^{\tiny(0)}/\text{tr}()$ for
$x=X_{0,d}$, $X_{0,b}^\pm$.  The arrows indicate times of photon
detection. Prototypical dot and laser parameters of $\Delta=2$ meV and
$\Omega_p^\pm=10$ $\mu$eV are used.  For the energy splitting $\delta$
between $X_{0,b}^\pm$ and $X_{0,d}$ we assume $\delta\ll T$, thus
obtaining the same scattering rates for $X_{0,b}^\pm\to X_{0,d}$ and
$X_{0,d}\to X_{0,b}^\pm$; we use $(2\Gamma^s)^{-1}=50$ ns and
$(2\Gamma^\gamma)^{-1}=0.86$ ns for the radiative decay, which
corresponds to the bulk-GaAs value.
}
\end{figure}

\begin{eqnarray}\label{eq:conditional}
  \rho_{t_n}^{\text{\tiny(0)}}&=&
        I\!\!P_0 U(t_n,t_{n-1})\dots
        I\!\!P_0 U(t_1,t_o)\rho_o\nonumber\\
	&&\qquad\times
	U^\dagger(t_1,t_o)I\!\!P_0\dots
	U^\dagger(t_n,t_{n-1})I\!\!P_0,
\end{eqnarray}    

\noindent where $U(t_n,t_{n-1})$ is the time evolution operator from
$t_{n-1}$ to $t_n$, and the superscript on $\rho^{\text{\tiny(0)}}$
indicates the requirement of no-photon emission. The quantity
$P_0(t)=\text{tr} (\rho_{t}^{\text{\tiny(0)}})$ then gives the
probability that, starting at time $t_o$ with $\rho_o$, no photons are
emitted up to time $t$; apparently, $P_0(t)$ is a monotonically
decreasing function since the probability of photon emission will
increase with time. If finally at time $t_n$ a photon is detected, we
know that the system has changed to the one-photon subspace
$\rho^{\text{\tiny(1)}}$ (we assume that $\Delta_t$ is short enough and
only {\em one}\/ photon is emitted per interval). What the photon
detector does, however, is to absorb the photon and to reset the
density matrix to \cite{hegerfeldt:93}:

\begin{equation}
  |0_\gamma\rangle\text{tr}_\gamma(
  I\!\!P_1 U(t_n,t_{n-1})\rho_{t_{n-1}}^{\text{\tiny(0)}}
  U^\dagger(t_n,t_{n-1})I\!\!P_1)\langle 0_\gamma|\;/\;\text{tr}(),
\end{equation}

\noindent with $I\!\!P_1={\bf 1}-I\!\!P_0$, and the denominator ensures
that $\text{tr}(\rho)=1$ after detection of $\gamma$.

In this work we apply the above scheme to the resonance luminescence of
exciton states in a single QD. Quite generally, the driving laser can
not only excite single excitons but also multi-exciton states (e.g.,
biexcitons). Thus, we use $x$ to denote generic multi-exciton states,
with $N_x=0,1,2,\dots$ the number of electron-hole pairs; the
corresponding dot Hamiltonian is of the form $\sum_x
E_x|x\rangle\langle x|$, with $E_x$ the multi-exciton energies. The
coupling to the pump laser with electric field ${\cal
E}_p^\sigma(t)={\cal E}_p^\sigma\cos(\omega_pt)$ is described within
the usual rotating-wave and dipole approximations, $-\frac 1 2
\sum_{\sigma,xx'} \Omega_p^\sigma(e^{ i\omega_pt}{\cal P}_{x'x}^\sigma+
e^{-i\omega_pt}{\cal P}_{xx'}^{\sigma\;*})|x'\rangle\langle x|$,
where:  $\sigma=\pm 1$ denotes left- and right-hand circular
polarization, respectively (we assume that, as in most semiconductors,
electron-hole pairs with given spin orientation can be selectively
created by circular polarized light); $\Omega_p^\sigma=\mu_o{\cal
E}_p^\sigma$ is the Rabi frequency with $\mu_o$ the dipole element of
the bulk semiconductor; ${\cal P}_{x'x}^\sigma=\int d{\bf r}\;\langle
x'| \hat\psi_{h,-\sigma}({\bf r})\hat\psi_{e,\sigma}({\bf
r})|x\rangle$, with $\hat\psi_{e,\sigma}^\dagger$
($\hat\psi_{h,\sigma}^\dagger$) creating an electron (hole) with spin
orientation $\sigma$. The spontaneous radiative decay of dot states is
accounted for by the Hamiltonian $\sum_{{\bf
k}\sigma,xx'}ig_k^\gamma(a_{{\bf k}\sigma}^\dagger {\cal
P}_{x'x}^\sigma-a_{{\bf k}\sigma}{\cal P}_{xx'}^{\sigma\;*})
|x'\rangle\langle x|$, where $g_k^\gamma=\mu_o(2\pi\omega_k)^{\frac 1
2}$, $a_{{\bf k}\sigma}^\dagger$ creates a photon with wavevector $\bf
k$ and circular polarization $\sigma$, and $\omega_k=ck/n$ is the
photon energy (with $c$ the speed of light and $n$ the semiconductor
refraction index).

Spin-flip scatterings in III-V semiconductors are conveniently
described as a two-step process \cite{khaetskii:00,sham:93}: First,
spin-orbit coupling introduces a small mixing of electron (hole) states
with different spin orientations; second, environment coupling (e.g.,
to phonons) mediates transitions between states with different spin
orientations. While in semiconductors of higher dimensionality all
different types of elastic and quasi-elastic scatterings contribute to
such spin-flip processes \cite{sham:93,potemski:99}, in semiconductor
QDs severe phase-space restrictions lead to a strong suppression, and
first experimental results indicate extremely long spin coherence times
\cite{gupta:99,kamada:99}. As a genuine model accounting for spin flips
in QDs we consider, within the spirit of Ref.~\cite{loss:98}, a
Caldeira-Leggett-type model where a set of harmonic oscillators
(assumed to be in thermal equilibrium) is coupled linearly to the
system spins by $\sum_j g_j^s\;{\bf S}\cdot({\bf b}_j+{\bf
b}_j^\dagger)/\sqrt 2$; here ${\bf b}_j$ are bosonic operators whose
free motion is governed by $\sum_j\omega_j^s\; {\bf
b}_j^\dagger\cdot{\bf b}_j$, and $g_j^s$ is the coupling constant to
the carriers in the dot.

Next, we assume that the time interval $\Delta_t$ of successive {\em
gedanken}\/ experiments is short as compared to the time evolution of
multi-exciton states due to the driving pump laser, but long as
compared to $\omega_p^{-1}$; thus, in the evaluation of $U$ of
Eq.~(\ref{eq:conditional}) we perform the Markov and adiabatic
approximations. Using the interaction representation according to
$\sum_x\Delta_x|x\rangle\langle x|$ (with
$\Delta_x=E_x-N_x\cdot\omega_p$), we then find within second order
perturbation theory for the time evolution of the conditional density
matrix $\rho^{\text{\tiny(0)}}$ \cite{hegerfeldt:93,hohenester:00}:

\begin{equation}\label{eq:rhodot}
  \dot\rho^{\text{\tiny(0)}}=
  -i(h_o\rho^{\text{\tiny(0)}}-\rho^{\text{\tiny(0)}}h_o^\dagger)+
  \hat{\cal J}^s\rho^{\text{\tiny(0)}},
\end{equation}

\noindent with $\dot\rho_{t_n}^{\text{\tiny(0)}}\equiv(
\rho_{t_n}^{\text{\tiny(0)}}-\rho_{t_{n-1}}^{\text{\tiny(0)}})/\Delta_t$,
the effective Hamiltonian $(h_o)_{xx'}=\delta_{xx'}\Delta_x-\frac 1 2
\sum_\sigma\Omega_p^\sigma({\cal P}_{xx'}^\sigma+{\cal
P}_{x'x}^{\sigma\;*})-i(\Gamma_{xx'}^\gamma+\Gamma_{xx'}^s)$, where
$\Gamma^{\gamma,s}$ are out-scattering contributions due to photon and
spin interactions, and $\hat{\cal J}^s$ accounts for spin-flip induced
in-scatterings \cite{remark.matrix} (note that because of our
restriction to the zero-photon subspace there are no corresponding
in-scatterings due to photon decay). Finally, whenever a photon is
detected the density matrix is reset to $\rho^{\text{\tiny(0)}}
\longrightarrow\hat{\cal J}^\gamma \rho^{\text{\tiny(0)}}$/\text{tr}(),
with the reset matrix $\hat{\cal J}^\gamma$ \cite{remark.matrix}.

\begin{figure}
\centerline{\includegraphics[width=0.75\columnwidth]{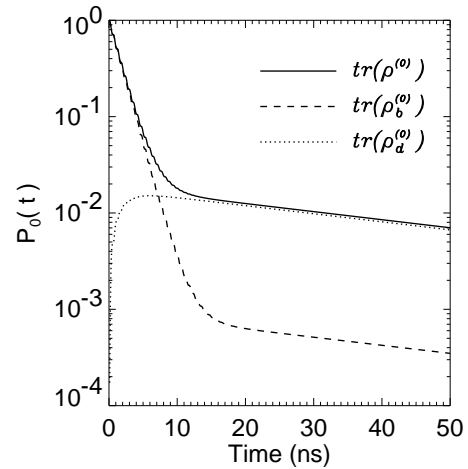}}
\caption{
Time-dependent probability $P_0(t)$ of no-photon emission for a
coherently driven system which is in the groundstate at time $0$ (same
parameters as in Fig.~2). The solid line corresponds to
$\text{tr}(\rho^{\tiny(0)})$, whereas the dotted and dashed lines show
the contributions of $\langle X_{o,d}|\rho^{\tiny(0)}|X_{o,d}\rangle$
and remainder.
}
\end{figure}

In our calculations we use prototypical dot parameters, which
correspond to the model structure presented in Ref.~\cite{troiani:00}
(see figure caption 2), and for the pump laser we assume linear
polarization (i.e., ${\cal E}_p^+={\cal E}_p^-$) and photon energy
$\omega_p$ tuned to the groundstate-exciton energy (see Fig. 1).
Because excited single-exciton states are energetically separated from
$E_{X_{o,b}^\pm}$ by the order of the confinement energy ($\sim$10--50
meV for most dots \cite{dot}), for moderate Rabi frequencies they do
not couple to the driving laser and are therefore neglected in our
analysis. As the only multi-exciton state we consider the biexciton
groundstate $B_o$, which approximately consists of
$X_{o,b}^++X_{o,b}^-$ (with other multi-exciton states again
energetically well separated). In Ref.~\cite{gupta:99} spin dephasing
times $\tau_s\sim3$ ns have been reported, which, however, have been
dominated by inhomogeneous broadening. Thus, as a representative value
within a single dot we here assume a value of 50 ns. Smaller values of
$\tau_s$ would not change the qualitative behavior of our results, as
long as they would be significantly longer than the radiative lifetime
$\sim$1 ns. Fig.~3 shows the probability $P_0(t)$ that up to time $t$
no photon has been emitted. At times below $\sim$10 ns the time
evolution is governed by the optical decay of the laser-excited exciton
states $X_{o,b}^\pm$; with increasing time, however, the probability of
a spin-flip process increases and becomes dominant at later times.
Because spin-forbidden transitions do not couple to the light, in this
regime $P_0(t)$ decreases only very slowly and the decay is determined
by additional spin-flip processes bringing back the system to the
bright sector.  We used these results to perform Monte-Carlo
simulations of resonance luminescence from a single QD (Fig.~2)
\cite{plenio:98}: At time $t_o$, we start with a density matrix
$\rho_o$; the time of the next-photon emission is then determined by
use of a random number $r\in [0,1)$ from $r=1-P_0(t)$ (in the numerical
solution of Eq.~(\ref{eq:rhodot}) we neglect the time discretization
$\Delta_t$); finally, after a photon count the density matrix is set
back to $\hat{\cal J}^\gamma \rho^{\tiny(0)}/tr()$, and the simulation
starts again.  Fig.~2(c) shows a single history of the occupation of
$X_{o,b}^\pm$: Rabi-type oscillations are interrupted by sudden jumps
(arrows), associated to the detection of emitted photons;
correspondingly, the dark-exciton contributions (Fig.~2(b)) slowly
increase in time, and are reset back to zero after photon detection
(note the beginning of a dark period at later times in Figs.~2(b,c)).
Finally, Fig.~2(a) shows the histogram of emitted photons: One clearly
observes bright and dark sectors, whose lengths are determined by
$\tau_s$. Experimental observation of single spin flips requires
coherent excitation of single QDs and efficent continuous broadband
counting of photons on a timescale of nanoseconds.


I am grateful to Elisa Molinari and Luis Vi\~na for a careful reading
of the manuscript and for most stimulating discussions. This work was
supported in part by the EU under the TMR Network "Ultrafast" and the
IST Project SQID, and by INFM through grant PRA-SSQI.

\end{multicols}


\begin{thebibliography}{10}
\bibitem{remark.detector} Experimentally, it is certainly impossible to
achieve complete 4$\pi$-coverage. However, as discussed in
Ref.~\cite{plenio:98}, smaller detector efficiency does not cause
changes of the qualitative behavior.
\end{thebibliography}
\end{document}